# Hydrogen as a Source of Flux Noise in SQUIDs

Zhe Wang[1,2,+], Hui Wang[2,+,*], Clare C. Yu[2,§] and R. Q. Wu[1,2,*,§]

[1]*State Key Laboratory of Surface Physics, Key Laboratory of Computational Physical Sciences, and Department of Physics, Fudan University, Shanghai 200433, China*

[2]*Department of Physics and Astronomy, University of California, Irvine, California 92617, USA*

+These authors contributed equally to this work.

§CCY and RQW are co-senior authors.

*Correspondence should be addressed to:huiw2@uci.edu,wur@uci.edu

Superconducting qubits are hampered by flux noise produced by surface spins from a varietyof microscopic sources. Recent experiments indicated that hydrogen (H) atoms may be one of those sources. Using density functional theory calculations, we report that Hatoms either embedded in, or adsorbed on,an$\alpha$-Al$_2$O$_3$(0001) surface have sizeable spin momentsranging from0.81 to 0.87$\mu_B$withenergy barriersforspin reorientation as low as ~10 mK. Furthermore, H adatoms on the surface attractgas molecules such as O$_2$, producing new spin sources. We propose coating the surface with grapheneto eliminate H-induced surface spins and to protect the surface from other adsorbates.

Superconducting circuitshave a wide variety of applications, e.g., photon detectors used in astrophysics[1],bolometers involved in dark matter searches [2], nanomechanical motion sensors [3], cavity quantum electrodynamics [4, 5], and quantum limited parametric amplifiers[6].However, their performance continues to be impaired by noise and dielectric loss produced by microscopic defects. While some progress has been made [7-9], identifying microscopic sources of noise remains a top priority. Of particular interest as a qubit is the superconducting quantum interference device (SQUID) [10]where a major problem is low-frequency 1/f flux noise generated by fluctuating spins residing on the surface of normal metals [11], superconductors [12, 13] and insulators [14].Proposed microscopic sources of spins have includedsurface spin clusters and correlated fluctuations [15, 16], electron spin exchange via the hyperfine interactions[17], and adsorbed OH or $O_2$ molecules [18, 19]. In particular, the suggestion of adsorbed $O_2$ molecules [19] has been supported by experimental measurements involving X-ray magnetic circular dichroism (XMCD) as well as measurements of susceptibility and flux noise [7]. Efforts to remove adsorbed $O_2$ molecules have significantly reduced the flux noise in SQUIDs, but have not completely eliminated it, implying that there are additional sources of flux noise[7]. Recent experiments have implicated hydrogen (H) atoms as a source of flux noise [8, 20] even though hydrogen is rarely associated with magnetism. Electron spin resonance (ESR) measurementsfind an energy splitting of ~1.42 GHz on sapphire (α-$Al_2O_3$(0001)) which is often used as a substrate and as a model of the native oxide layer on Al SQUIDs.1.42 GHz coincides with the hyperfine splitting of a free H atom. To explain this observation and to find ways to eliminate magnetic noise in Al SQUIDs, we investigated the magnetic states of different arrangements of H atoms in and on the surface of aluminum oxides.

In this work, we useddensity functional theory (DFT) to investigateHatoms as a source of flux noise on α-$Al_2O_3$(0001). H atoms canoccupy interstitial sites in the bulk sapphire or be adsorbedon various surface sites. In either case they can produce a sizeable local magneticmoment. Furthermore, H atoms on α-$Al_2O_3$(0001) facilitate the adsorption of other molecules such as $O_2$ that can produce additionalfluctuating spins. The binding energies of H adatoms and H+$O_2$ co-adsorbates are large and hence cannot be easily removed through heating. We suggest that the flux noise from H atoms can be reduced by coating the α-$Al_2O_3$(0001) surface with grapheneto remove unpaired electrons from H/α-$Al_2O_3$(0001) and prevent other magnetic species from being adsorbed.

Our DFT calculations used the projector augmented wave method (PAW) implemented in the Vienna *ab initio* simulation package (VASP) [21, 22]. Exchange-correlation interactions were included using the generalized-gradient approximation (GGA) with the Perdew-Burke-Ernzerhof (PBE) functional[23]. The α-$Al_2O_3$(0001) surface was mimicked by building a slab model that consists of 18 atomic layers and a vacuum gap about 15 Å thick to avoid spurious interaction. A 3×3×1 Monkhorst-Pack mesh [24]was used to sample the Brillouin zone to optimize the 2x2 supercell with criteria that force acting on each atom was less than 0.01 eV/Å. The van der Waals correction was implemented using the PBE-optB86b functional [25]. The energy cutoff for the plane-wave expansion was set to 600 eV, as in ourprevious studiesof H[26, 27].For direct comparison with experiment, the X-ray absorption spectroscopy(XAS) and XMCD spectra, as well as the ESR frequencies were calculated using the full potential linearized augmented plane-wave (FLAPW) method[28, 29].To identify plausible sources of 1/f noise, we calculated themagnetic anisotropy energy (MAE) which is the energy barrier for spin rotation. To determine the MAEat the micro-electron volt (μeV) level, we usedtorque methods[30]that evaluate the expectation values of angular derivatives of the Hamiltonian with respect to the polar angle $\theta$ and azimuthal angle $\phi$ of the spin moment, i.e., $\tau(\theta) = \frac{\partial E_{total}(\theta)}{\partial \theta} = \sum_{occ} \langle \psi_{i,k} | \frac{\partial H_{SO}}{\partial \theta} | \psi_{i,k} \rangle$ ,as in studies of magnetic molecules and magnetostrictive alloys[31, 32].

Adsorbed hydrogen comes from atmospheric $H_2$ or $H_2O$ molecules. So we examined the adsorption and dissociation of $H_2$ and $H_2O$ molecules on α-$Al_2O_3$(0001) and found that $H_2$ binds weakly (binding energy ~ -0.14 eV) while $H_2O$ binds strongly (binding energy ~ -1.15 eV) to the α-$Al_2O_3$(0001) surface. *Ab initio* molecular dynamics (AIMD) simulations demonstrate that $H_2$ can be easily desorbed from the surface whereas $H_2O$ tends to disassociate into OH and H (see Fig. S1 and S2 in supplementary materials), consistent with previous reports[33]. Al samples and their thin native oxide layers likely contain a small amount of atomic H under ambient conditions[34-36]. As depicted in Fig. 1(a), atomic H can be easily trapped in cage-like interstitial sites in α-$Al_2O_3$.According to our Climbing Image-Nudged Elastic Band (CI-NEB) simulations [37],the energy barrier for an H atom diffusing from the interior along the path indicated in Fig. 1(a) is as high as ~1.07eV(Fig. 1(b)). Our AIMD simulations at 300 K demonstrate that H atoms do not drift away from a cage deep inside bulk sapphire over a period of 4 picoseconds (see Fig. S3(a)). Thus, H atoms (denoted $H_{inters}$) in Al SQUIDs may be impurities in interstitial sites inside the oxide layer and adsorbates on the surfaces. (Note that H is non-magnetic inmetallic Al.) In the following discussionswe will focus on the energetic and magnetic properties of interstitial H atoms embedded in different layers of bulk α-$Al_2O_3$as well as adsorbates on the surface.

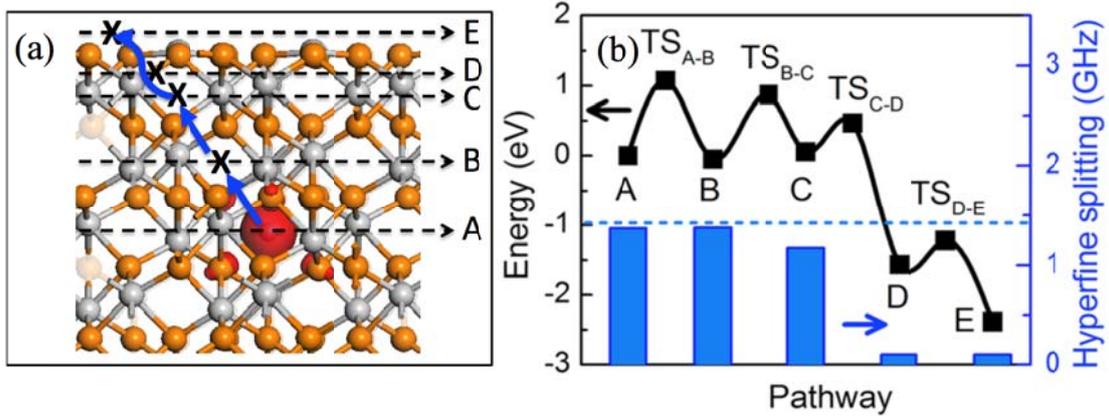

Fig. 1. (color online) (a) Left panel shows the geometries and spin densityof an Hatom embeddedin α-$Al_2O_3$(0001).The grey, orange and green balls represent Al, O and H atoms, respectively.The spin density (of majority spins) of embedded H atomsare represented by redisosurfaces (0.05 e/Å$^3$).Black crossesshow the positions along the diffusion path (blue arrows) for embedded H atoms heading towards the surface, with A, B, C, D and E denoting the interstitial sites in different layers. (b) Left axis shows the relative total energy as anH atom diffusesfrominterior sites to the surface. Energies at $TS_{A-B}$,$TS_{B-C}$,$TS_{C-D}$ and$TS_{D-E}$ indicate the diffusion barriers between two adjacent interstitial sites. Right axis shows the calculated ESR values corresponding to each interstitial site. The horizontal blue dashed line represents the experimental ESR value [8].

$H_{inters}$ hardly interacts with adjacent atoms, thusretaining its atomic properties.Fig 1(a) shows the large spin densityaround $H_{inters}$ with a moment ~0.87 $\mu_B$. Calculations with large unit cells find antiferromagnetic (AFM) interactions between that $H_{inters}$ atoms in α-$Al_2O_3$(0001), with exchange energies of -0.12meV (~1.4 K) when the separation between two $H_{inters}$ is 4.8 Å, and -0.03meV (~0.4 K) for a separation of 9.6 Å (see Table 1). The MAE of $H_{inters}$ is smaller than 1 μeV (< 10 mK) which is almost beyond the limit of DFT approaches, indicating that the spin orientation energy is virtually isotropic. According to our previous Monte Carlo simulations of classical anisotropic XY spins[19], this implies that $H_{inters}$ atoms can produce 1/f flux noise.

Fortunately, the native oxide layeron Alis typically very thin and $H_{inters}$ atoms are likely to be driven to the surface by the large energy difference between the bulk and the surface as shown in Fig. 1(b).Energy barriers gradually decrease as $H_{inters}$ moves towards the surface of α-$Al_2O_3$(0001). AIMD simulations of $H_{inters}$ atoms embedded in interstitial sites near the surface (layer C in Fig. 1(a)) demonstrated that they drift to the oxygen site on the α-$Al_2O_3$(0001) surface within 5 picoseconds at 600 K which is consistent with experiment [8](see Fig. S3(b)). Therefore, the apparent density of $H_{inters}$ should be lowunder ambient conditions. However, a recent experiment on a thick sapphire sample by de Graaf*et al.*[8] found a strong ESR signal at ~1.42 GHz, indicating a rather high density of atomic H (~$2.2\times10^{17}$ $m^{-2}$). Our calculations found that the ESR hyperfine splitting for $H_{inters}$ atoms embedded in different layers of sapphire is between 1.28 and ~1.36 GHz(see Fig. 1(b)), very close to the experimental result of de Graaf*et al.*[8]. A peak in the flux noise of an Al/sapphire fluxmonqubit at ~1.4 GHz was also reported by Quintana *et al.* [20], which could be caused by the spin fluctuations of interstitial H atoms. Therefore, $H_{inters}$ atoms could produce flux noise that could be reduced by annealing at high temperatures[8].

Since both the outward segregation of $H_{inters}$ and the dissociation of $H_2O$may result in H atoms on the α-$Al_2O_3$(0001) surface, we found the preferred adsorption sites and binding energies ofan H adatomusing:

$$E_b = E_{H/Al_2O_3(0001)} - E_{Al_2O_3(0001)} - E_H \quad (1)$$

$E_{H/Al_2O_3(0001)}$ and $E_{Al_2O_3(0001)}$ are the total energies of the α-$Al_2O_3$(0001) slab with and without an H atom, respectively. $E_H$ is the total energy of the free H atom. By consideringan H atom adsorbed on top of O, Al, and O-O bridge sites, we found that the most stable site is on top of the oxygen atom on the α-$Al_2O_3$(0001) surface (denoted as "$H_{atop-O}$"),(see Fig.2(a)). The binding energy and bond length of H-O are about-1.07 eV and 0.98 Å, respectively. Another stable but less desirable adsorption site for H is on top of the surface Al site (denoted as "$H_{atop-Al}$") (see Fig. 2(b)) with anH-Al bond length of 1.67 Å and a binding energy of -0.39 eV. The energy barrier is ~0.26 eVfor the conversion from $H_{atop-Al}$ to $H_{atop-O}$and is 0.94 eVin the reverse process(seeFig. 2(c)). From these numbersthe $H_{atop-Al}$ geometryoccurs much less frequently than $H_{atop-O}$ for H adatoms on the α-$Al_2O_3$(0001) surface.

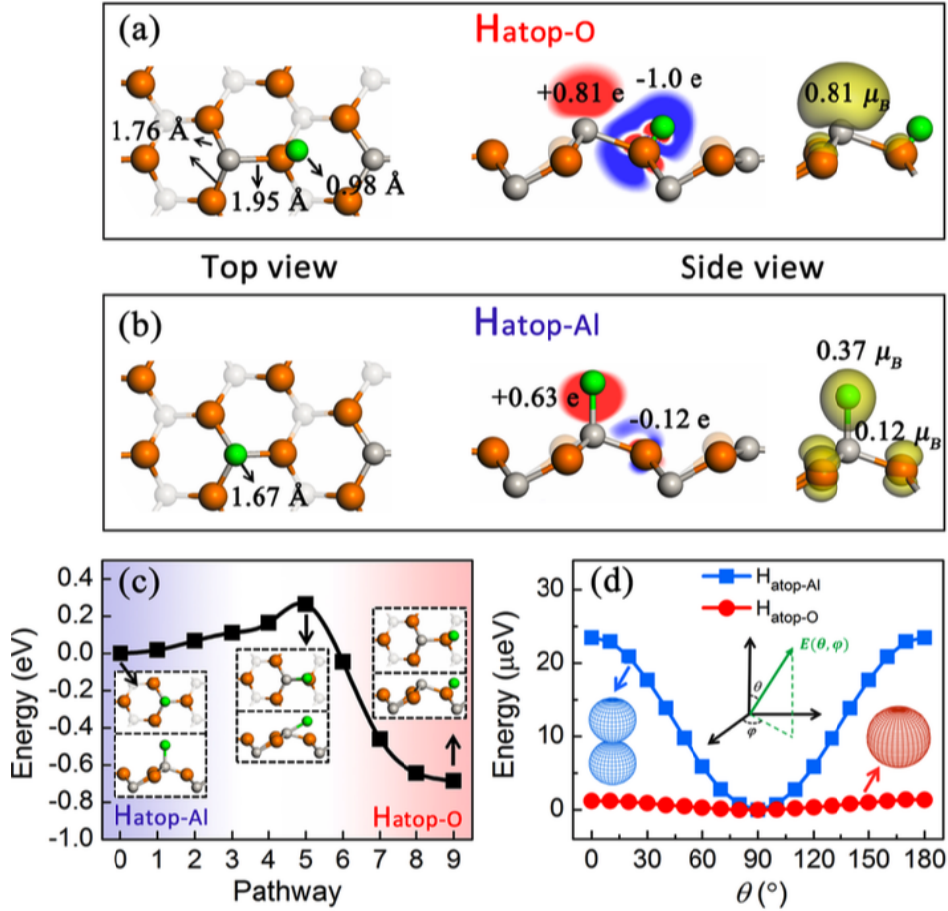

*Fig. 2 (color online) (a) Schematic geometries of an H atom adsorbed on the O site of an α-Al$_2$O$_3$(0001) surface. Only atoms near the adsorption site are shown. Al, O, H atoms are colored as in Fig. 1. Bond lengths are expressed in angstroms. Charge depletion and accumulation are represented by blue and red, respectively. The lime green isosurface depicts the distribution of spin density. (b) The same as (a) but with an H atom adsorbed on the Al site of an α-Al$_2$O$_3$(0001) surface. (c) Reaction pathway of the H adatom hopping from the H$_{atop-Al}$ site to the H$_{atop-O}$ site. The horizontal dashed lines indicate the energy barrier. Insets show the top and side view of atomic arrangements corresponding to different states. (d). Relative total energy versus the polar angle θ of the spin direction with respect to the surface normal for the H$_{atop-O}$ (red line) and H$_{atop-Al}$ geometries (blue line) on an α-Al$_2$O$_3$(0001) surface. The left and right insets show the isoenergy surfaces of the MAE versus the polar and azimuthal angles that are sketched in the central inset.*

Our Bader charge analysis indicates that the H$_{atop-O}$ adatom donates its charge to the adjacent O atoms (0.16 e) and to the neighboring Al atom (0.81 e), as depicted by the charge redistribution in Fig. 2(a). As a result, the topmost Al atom is strongly magnetized with a spin moment of ~0.81μ$_B$, with a spin density distribution shown in Fig. 2(a). In contrast, H$_{atop-Al}$ gains electrons from the Al atom underneath it and the three neighboring O atoms (see the charge redistribution in Fig.2(b)). This results in magnetic moments of 0.37μ$_B$ and 0.12μ$_B$ for the H atom and each of the three surface O atoms, respectively. As shown in Fig. 2(d), the MAE is almost isotropic for H$_{atop-O}$, implyingeasy spin fluctuations in every direction. For $H_{atop-Al}$, the calculated MAEbetween the spin orientation in and out of the surface plane is about -24μeV, showing that the easy axis lies in the surface plane. However, the energy barrier to spin rotation in the surface plane is extremely small [~ 1μeV or 10mK].

*Table 1. Calculated exchange interaction energies, commonly denoted by J, at different separations for $H_{inters}$,$H_{atop-O}$, $H_{atop-Al}$, $H_{atop-O}+O_2$ and $O_2$ molecules in or on α-$Al_2O_3$(0001). The data for $O_2$ molecules comes from previous studies [19]. Positive values correspond to ferromagnetic interactions and negative values to antiferromagnetic interactions.*

|  | 4.8 Å | 9.6 Å |
|---|---|---|
| $H_{inters}$ (this work) | -0.12meV (1.4 K) | -0.03meV (0.4 K) |
| $H_{atop-O}$ (this work) | -5.05meV(60.6 K) | -0.01meV(0.1 K) |
| $H_{atop-Al}$ (this work) | 0.73meV(8.8 K) | 0.02meV(0.2 K) |
| $H_{atop-O}+O_2$ (this work) | -0.17meV (2.0 K) | -0.1μeV (~0 K) |
| $O_2$ molecule | 0.14meV (1.7 K) | 0.05meV (0.6 K) |

The noise spectrum depends on spin-spin interactions. As shown in Table 1, our DFT calculations with 2×2 and 4×4 supercells indicate $H_{atop-O}$ atoms interact antiferromagnetically (AFM) on α-$Al_2O_3$(0001), with exchange energies of -5.05meV (~60.6 K) when the separation between two $H_{atop-O}$ is 4.8 Å, and -0.01meV (~0.1 K) for a separation of 9.6 Å. In contrast, the $H_{atop-Al}$ induced magnetic moments interact ferromagnetically (FM), with exchange energies of 0.73meV (~8.8 K) when two $H_{atop-Al}$ atoms have a separation of 4.8 Å, and 0.02meV (~0.2 K) for a separation of 9.6 Å. Together with the small MAE discussed above, both $H_{atop-O}$ and $H_{atop-Al}$ could produce 1/f magnetic flux noise.

WhichH configuration dominates the flux noise on α-$Al_2O_3$(0001)?From the energetics in Fig. 1 and Fig. 2 for H segregation and adsorption, we find that the order of apparent densities (n) of H atoms in or on α-$Al_2O_3$(0001) is: n($H_{atop-O}$) > n($H_{inters}$) > n($H_{atop-Al}$). Our ESR calculations of the hyperfine splitting for $H_{atop-O}$is essentially zero, due to the complete depletion of its charge. The hyperfine splitting for $H_{atop-Al}$is 0.53 GHz, but this was not seen experimentally, consistent with our estimate of its small concentration.The surface to volume ratio implies that the ESR measurements[8]are dominated by the much more numerous H atoms embedded in the thick sapphire bulk, rather than by the surface spins.

Although $H_{atop-O}$ by itself is not magnetic, we found that $H_{atop-O}$ adatoms can attract other molecules from the atmosphere to the surface. In previous studies, we identified $O_2$ molecules as a possible source of 1/f noise [19], but these can either be removed by raising the temperature above 50 K due to the small binding energy (~ -0.15 eV per molecule) or avoided by protecting the surface with molecules that have a higher binding energy such as ammonia [7, 19]. In the presence of $H_{atop-O}$, the binding energy of an $O_2$ molecule next to an H adatom increases to around -2.9 eV, mainly due to significant charge rearrangement. In the most stable geometry, the $O_2$ bond lies almost parallel to the α-$Al_2O_3$(0001) surface as shown in Fig. 3(a), and gains a charge of +1.0e from the surrounding Al atoms to become "$O_2^-$". The O-O bond length stretches by 16%, which is very different from the adsorption of an $O_2$ molecule on a bare α-$Al_2O_3$(0001) surface. The calculated magnetic moment of the $H_{atop-O}+O_2^-$ complex is 1.0 $\mu_B$, with an easy axis along the O-O bond and an MAE of ~26 μeV (~0.30 K). This magnetic complex is a possible noise source and should form easily if $H_{atop-O}$ is present.

Note that de Graaf *et al.* suggested $O_2^-$ as the possible source of the central peak in their ESR experiment [8], but there are a number of possibilities since g=2.0 is characteristic of many spin systems. One way to experimentally confirm our prediction of $H_{atop-O}+O_2$ on α-$Al_2O_3$(0001) would be with XAS and XMCD spectra. According to our DFT calculations, the energies of the two $\pi_{2p}^*$ states of $O_2$ are split into two as an additional electron is transferred from an Al atom to the $O_2$ in the $H_{atop-O}+O_2$ complex as shown in Fig. 3(b). In the unoccupied branch, $\pi_{m=1}^*$ and $\pi_{m=-1}^*$ have different weights because of the joint effect of magnetization and spin orbit coupling. The selection rules for dipole transitions ensure that left-circularly polarized light (LCPL) excites electrons from 1*s* core states to the unoccupied $\pi_{2p(m=1)}^*$ states, whereas right-circularly polarized light (RCPL) excites electrons from 1*s* to $\pi_{2p(m=-1)}^*$ states. The imbalance between ±m in the unoccupied states produces an XMCD peak at the onset of the k-edge of $O_2^-$ as seen in Fig. 3(c). The XAS has more features in the higher energy region due to transitions to other orbitals.

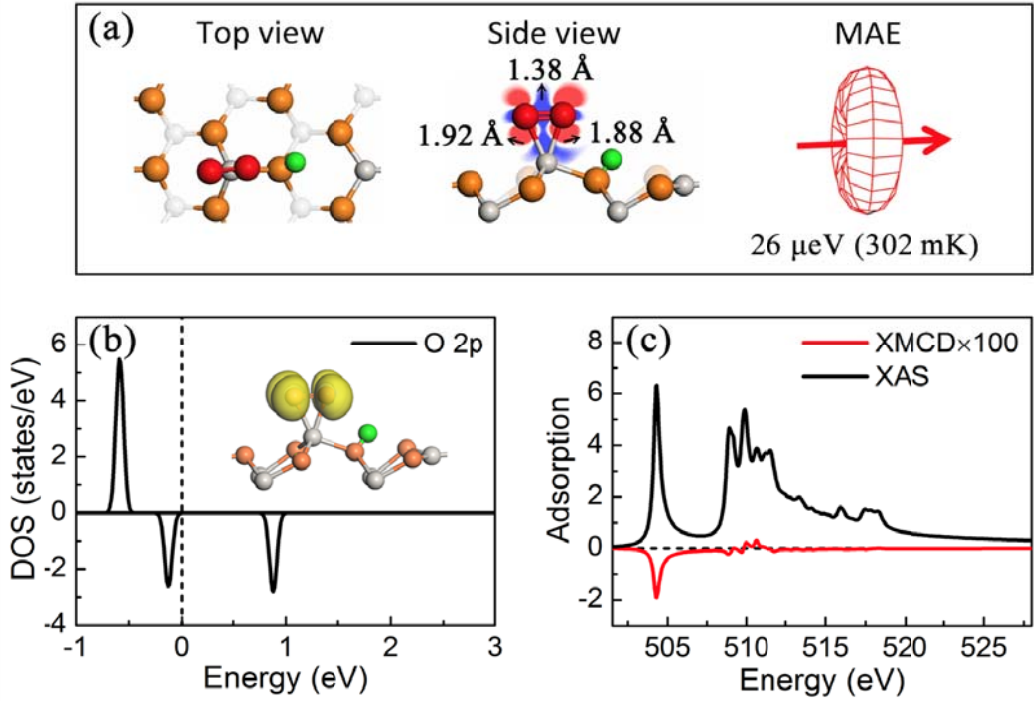

*Fig. 3 (color online) (a) The atomic geometry and charge redistribution of an $O_2$ molecule adsorbed on $H_{atop-O}/\alpha$-$Al_2O_3(0001)$. Al, O, H atoms are colored as in Fig. 1. $E_{tot}(\theta, \varphi)$ is given in the right figure, with an arrow indicating the easy axis. Charge depletion and accumulation is represented by blue and red colors, respectively. (b) The PDOS of $O_2$ molecules adsorbed on $H_{atop-O}/\alpha$-$Al_2O_3(0001)$. The inset gives the isosurface of the spin density. (c). Calculated XAS and XMCD spectra of the oxygen K-edge for $O_2$ molecules associated with $H_{atop-O}/\alpha$-$Al_2O_3(0001)$.*

Since H atoms either embedded in $\alpha$-$Al_2O_3(0001)$ or adsorbed on its surface can produce flux noise, we need to find ways to remove them and diminish their magnetic moments. The binding energy of $H_{atop-O}$ is rather large so it is difficult to completely eliminate them by annealing. We propose using graphene as a protective coating due to its high structural stability and electron affinity to further 1) reduce the $H_{atop-O}$ induced magnetization through charge transfer to the graphene; and 2) prevent $H_2O$, $O_2$ and other molecules from reaching the surface. Graphene has a small lattice mismatch (~1%) with $\alpha$-$Al_2O_3(0001)$, and our calculations indicate that it binds strongly to $H_{atop-O}/\alpha$-$Al_2O_3(0001)$, with a binding energy of -0.65 eV/per unit cell. Significant charge transfer to graphene occurs as shown by the charge density difference in Fig. 4(a) and the z-dependence of its planar-average in the out-of-plane direction in Fig. 4(b), as well as by the downshift of graphene bands in Fig. 4(c). The $p_z$ orbitals of carbon absorb electrons from the underlying Al atoms, resulting in charge being redistributed into delocalized electrons in graphene. As a result, the magnetic moment of $H_{atop-O}/\alpha$-$Al_2O_3(0001)$ is completely quenched. By preventing other gas molecules from adsorbing, agraphene coatingcould effectively reduce flux noise.Another issue is whether $H_{atop-O}$ diffuses across graphene and further recovers the magnetism.One can see that the magnetism is gradually recovered after H atoms diffuse across andadsorb on graphene, which induces aspin-polarized state that is essentially localized on the carbon sublattice[right axis in Fig. 4(d) and Fig. 4(e)], in good agreement with recent report of controlling graphene magnetism by using H atoms[38].However, the diffusive energy barrier is as high as 5.1eV when an H atom passes through graphene on α-$Al_2O_3$(0001) [Fig. 4(d)], indicating the diffusive behavior of H atoms is completely suppressed once they chemisorbed on $H_{atop-O}$ site.

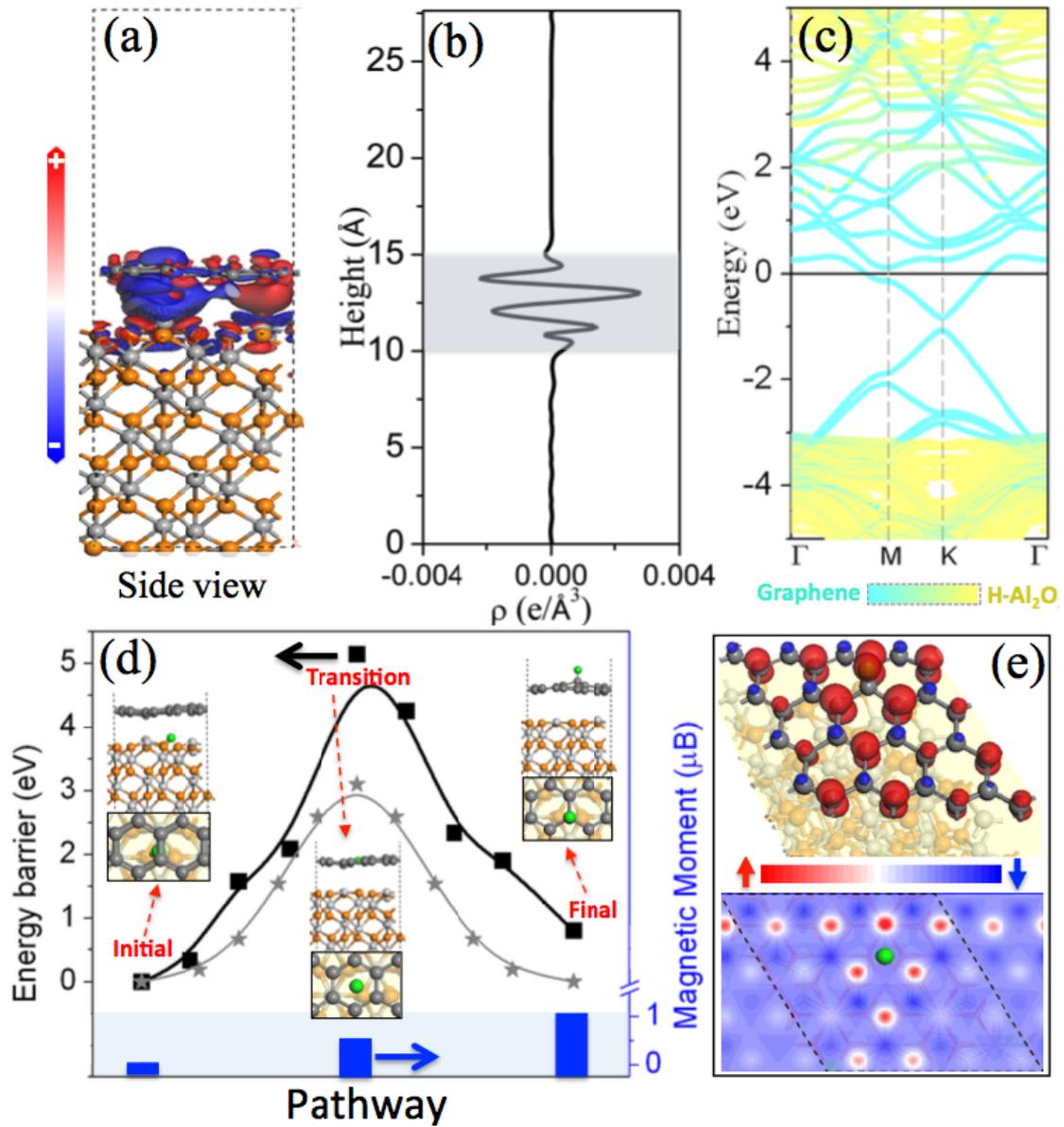

Fig. 4 (color online) (a) Atomic geometryand charge redistribution of graphene/$H_{atop-O}$/α-$Al_2O_3$(0001).Color coding is the same as Fig. 1.(b) The planar-average of charge density difference along the surface normal. Shaded area corresponds to the region between graphene and the$H_{atop-O}$/α-$Al_2O_3$(0001) substrate where the most charge is

*transferred.(c) Electronic band structure of graphene/$H_{atop-O}$/α-$Al_2O_3$(0001) with the color bar indicating the relative weights of graphene (light blue) and the $H_{atop-O}$/α-$Al_2O_3$(0001) substrate (light yellow). Horizontal black line represents the Fermi level. (d) The relative total energy as an H atom diffuses from $H_{atop-O}$ site on α-$Al_2O_3$(0001) surface to the $H_{atop-C}$ site on graphene sheet. Grey solid line represents H atom diffuse through freestanding graphene. Insets are the top and side view of atomic configurations corresponding to different state. Right axis shows the calculated magnetic moment corresponding to each state. The grey, orange, green and dark grey balls represent Al, O H and C atoms, respectively. (e) Spin density distribution in three (upper panel) and two dimension (lower panel) for H atom adsorbed on graphene sheet. Red and blue represent majority and minority spin channel, respectively.*

In summary our systematic DFT calculations demonstrate that H atoms embedded in ($H_{inters}$) or adsorbed on ($H_{atop-O}$) α-$Al_2O_3$(0001) have sizeable magnetic moments that can produce 1/f flux noise, owing to their small MAEs (a few mK) and moderate exchange interactions. In addition,$H_{atop-O}$may also strongly attract gas molecules from the environment, resulting in additional sources of flux noise. We propose coatingAl SQUIDs with a layer of graphenethatwould not only protect the surface from other gas molecules, but also eliminate the magnetism produced by adsorbed H atoms. Our studies provide insights and strategies for reducingsources of magnetic noiseinsuperconducting circuits.

**Acknowledgement:**

Work at UCI was supported by DOE-BES (HW and RQW, Grant No. DE-FG02-05ER46237). Work at Fudan(ZW) was supported by the National Science Foundation of China under Grant No. 11474056 and National Basic Research Program of China under grant number 2015CB921400. CCY was supported in part by a gift from Google (Google Gift No. 2502) and a grant from the UC Office of the President Multicampus Research Programs and Initiatives (MRP-17-454755). Computer simulations were performed at the U.S. Department of Energy Supercomputer Facility (NERSC).

**Reference:**

[1]    P. K. Day, H. G. LeDuc, B. A. Mazin, A. Vayonakis, and J. Zmuidzinas, Nature **425**, 817 (2003).

[2]    A. Fleischmann, C. Enss, and G. M. Seidel, in *Cryogenic particle detection* (Springer-Verlag, Berlin, 2005).

[3]    C. A. Regal, J. D. Teufel, and K. W. Lehnert, Nat. Phys. **4**, 555 (2008).

[4]    X. Mi, J. V. Cady, D. M. Zajac, P. W. Deelman, and J. R. Petta, Science **355**, 156 (2017).

[5]    A. Stockklauser *et al.*, Phys. Rev. X **7**, 011030 (2017).

[6]     M. Hatridge, R. Vijay, D. H. Slichter, J. Clarke, and I. Siddiqi, Phys. Rev. B **83**, 134501 (2011).
[7]     P. Kumar *et al.*, Phys. Rev. Applied **6**, 041001 (2016).
[8]     S. E. de Graaf, A. A. Adamyan, T. Lindström, D. Erts, S. E. Kubatkin, A. Y. Tzalenchuk, and A. V. Danilov, Phys. Rev. Lett. **118**, 057703 (2017).
[9]     S. E. de Graaf, L. Faoro, J. Burnett, A. Adamyan, A. Y. Tzalenchuk, S. Kubatkin, T. Lindström, and A. Danilov, arXiv:1705.09158  (2017).
[10]    B. G. Levi, Phys. Today **62**, 14 (2009).
[11]    H. Bluhm, N. C. Koshnick, J. A. Bert, M. E. Huber, and K. A. Moler, Phys. Rev. Lett. **102**, 136802 (2009).
[12]    R. H. Koch, D. P. DiVincenzo, and J. Clarke, Phys. Rev. Lett. **98**, 267003 (2007).
[13]    S. Sendelbach, D. Hover, A. Kittel, M. Mück, J. M. Martinis, and R. McDermott, Phys. Rev. Lett. **100**, 227006 (2008).
[14]    H. Bluhm, J. A. Bert, N. C. Koshnick, M. E. Huber, and K. A. Moler, Phys. Rev. Lett. **103**, 026805 (2009).
[15]    S. Anton *et al.*, Phys. Rev. Lett. **110**, 147002 (2013).
[16]    S. Sendelbach, D. Hover, M. Mück, and R. McDermott, Phys. Rev. Lett. **103**, 117001 (2009).
[17]    J. Wu and C. C. Yu, Phys. Rev. Lett. **108**, 247001 (2012).
[18]    D. Lee, J. L. DuBois, and V. Lordi, Phys. Rev. Lett. **112**, 017001 (2014).
[19]    H. Wang, C. Shi, J. Hu, S. Han, C. C. Yu, and R. Q. Wu, Phys. Rev. Lett. **115**, 077002 (2015).
[20]    C. M. Quintana *et al.*, Phys. Rev. Lett. **118**, 057702 (2017).
[21]    G. Kresse and J. Furthmuller, Phys. Rev. B **54**, 11169 (1996).
[22]    G. Kresse and J. Hafner, Phys. Rev. B **49**, 14251 (1994).
[23]    J. P. Perdew, K. Burke, and M. Ernzerhof, Phys. Rev. Lett. **77**, 3865 (1996).
[24]    H. J. Monkhorst and J. D. Pack, Phys. Rev. B **13**, 5188 (1976).
[25]    S. Grimme, J. Antony, S. Ehrlich, and H. Krieg, J. Chem. Phys. **132**, 154104 (2010).
[26]    H. Wang, S. Li, H. He, A. Yu, F. Toledo, Z. Han, W. Ho, and R. Q. Wu, J. Phys. Chem. Lett. **6**, 3453 (2015).
[27]    S. W. Li, A. Yu, F. Toledo, Z. M. Han, H. Wang, H. Y. He, R. Q. Wu, and W. Ho, Phys. Rev. Lett. **111**, 146102 (2013).
[28]    R. Q. Wu and A. J. Freeman, J. Magn. Magn. Mater. **200**, 498 (1999).
[29]    E. Wimmer, H. Krakauer, M. Weinert, and A. J. Freeman, Phys. Rev. B **24**, 864 (1981).
[30]    X. D. Wang, R. Q. Wu, D. S. Wang, and A. J. Freeman, Phys. Rev. B **54**, 61 (1996).
[31]    J. Hu and R. Q. Wu, Phys. Rev. Lett. **110**, 097202 (2013).
[32]    H. Wang, Y. N. Zhang, R. Q. Wu, L. Z. Sun, D. S. Xu, and Z. D. Zhang, Sci. Rep. **3**, 3521 (2013).
[33]    K. C. Hass, W. F. Schneider, A. Curioni, and W. Andreoni, Science **282**, 265 (1998).
[34]    L. Gordon, H. Abu-Farsakh, A. Janotti, and C. G. Van de Walle, Sci. Rep. **4**, 7590 (2014).
[35]    Y.-H. Lu and H.-T. Chen, Phys. Chem. Chem. Phys. **17**, 6834 (2015).
[36]    A. B. Belonoshko, A. Rosengren, Q. Dong, G. Hultquist, and C. Leygraf, Phys. Rev. B **69**, 024302 (2004).
[37]    G. Henkelman, B. P. Uberuaga, and H. Jónsson, J. Chem. Phys. **113**, 9901 (2000).
[38]    H. González-Herrero *et al.*, Science **352**, 437 (2016).